\documentclass[11pt]{article}
\usepackage{amssymb}
\usepackage{amsfonts}
\usepackage{graphics}
\begin{document}
\begin{titlepage}

September 2003 \hfill{UTAS-PHYS-2003-06}\\
\vskip 0.6in
\begin{center}
{\Large {\bf On the Standard Model and Strongly-Correlated Electron Systems}}
\\[5pt]
\end{center}

\normalsize
\vskip .4in

\begin{center}
J. D. Bashford, \\
{\it School of Mathematics and Physics, University of Tasmania} \\
{\it PO Box 252-37, Hobart 7001, Tasmania Australia} \\
jdb@postoffice.utas.edu.au \\

\end{center}
\par \vskip 1cm

\begin{center}
{\Large {\bf Abstract}}\\
\end{center}
Higlighting certain similarities between the two-dimensional Luttinger
liquid model and the effective fermionic theory obtained from the hypercharge
Lagrangian, we argue the case for a new type of Standard Model extension.

\vspace{1cm}


\end{titlepage}
\section{Introduction}	
The fields of condensed matter and quantum field theory have a long history
of cross-pollination. Recently Volovik \cite{vvkbk} has written extensively on 
similarities between the electroweak sector of the Standard Model (SM) and the 
superfluid $^{3}$HeB and it
seems logical to seek quantum field-theoretic analogues of recent developments in condensed matter. 
In this letter we argue that phenomena such as chiral surface states in unconventional superconductors e.g. 
Sr$_{2}$RuO$_{4}$ \cite{Luke,Morinari} and 
Bi$_{1-x}$Ca$_{x}$MnO$_{3}$ \cite{Yoon}, 
chiral superfluidity in $^{3}$HeA \cite{vvkbk,Goryo1} or the slave-boson
technique \cite{Andrei,voit} may well have relevance to the
non-perturbative sector of the Standard Model.

Recently we have learned that Terazawa \cite{Terazawa} observed, 
in the context of minimal supersymmetric left-right SM extensions, that the 
16 chiral fermions of a single family in the SM naturally decompose by a 
slave-boson approach into  four fermions and four bosons. Moreover 
this scheme has been extended \cite{Volovik1}, \cite{Volovik2} to include a 
gauged family symmetry.
In fact our model \cite{me}, which has developed independently from the 
former, has the same basic ingredients: a restoration of left-right symmetry 
plus modification of the colour $SU(3)$ to produce fermion families.

Our motivation begins by noting the four-fermi terms obtained upon 
integrating the hypercharge boson from the $U(1)$ sector of the SM:
\begin{eqnarray}
{\cal L}_{hyp}&=&\bar{\psi} i \gamma^{\mu}\partial_{\mu} \psi 
+c^{2}_{L}(\bar{\psi}_{L}\lambda^{\alpha}\gamma^{\mu}\psi_{L})^{2}+
 c^{2}_{R}(\bar{\psi}_{R}\lambda^{\alpha}\gamma^{\mu}\psi_{R})^{2} \nonumber \\
& &\mbox{}+ c_{R}c_{L} (\bar{\psi_{L}}\lambda^{\alpha}\gamma^{\mu}\psi_{L}
\bar{\psi}_{R}\lambda_{\alpha}\gamma_{\mu}\psi_{R}+
\bar{\psi_{R}}\lambda^{\alpha}\gamma^{\mu}\psi_{R}
\bar{\psi}_{L}\lambda_{\alpha}\gamma_{\mu}\psi_{L}) \label{hyp}
\end{eqnarray}
Here $c_{L,R}$ denote the SM couplings while colour and
isospin degrees of freedom are denoted generically by unitary matrices
$\lambda^{a}$. We wish to compare this with the two-dimensional Luttinger liquid Lagrangian \cite{voit,Haldane}, 
\begin{equation}
{\cal L}_{Lut}=\bar{\psi} \gamma^{\mu}\partial_{\mu} \psi + g_{2}\bar{\psi_{L}}\lambda^{\alpha}\gamma^{\mu}
\psi_{L}\bar{\psi}_{R}\lambda_{\alpha}\gamma_{\mu}\psi_{R}
+g_{4}((\bar{\psi}_{L}\lambda^{\alpha}\gamma^{\mu}\psi_{L})^{2}+
(\bar{\psi}_{R}\lambda^{\alpha}\gamma^{\mu}\psi_{R})^{2}), \label{lut1}
\end{equation}
where now the unitary matrices $\lambda$ denote an internal flavour symmetry.

If one identifies the terms $\bar{\psi}_{L,R}\gamma^{\mu}\psi_{L,R}$ with 
left-and right-moving charge density operators, which in the Luttinger
liquid are of the form $\psi_{L,R}\psi_{L,R}$, i.e.  fermion-fermion condensates, then there is a, superficial at least, resemblance between the two:

The latter model has, in two dimensions, a number of remarkable features.
The cross-term $g_{2}$ modifies the pole structure of the fermion propagator.
Note that the Fierz re-ordering of the $c_{L}c_{R}$ term in the four-dimensional
theory, Eq.(\ref{hyp}), also leads to an NJL-type four-fermi interaction, dynamically
breaking chiral symmetry.

The $g_{4}$ term in Lagrangian (\ref{lut1}) lifts any residual degeneracies, similar to a 
hopping matrix element between spin chains \cite{voit}, and leads in two 
dimensions to the seperation of fermionic degrees of freedom into
charge and spin fluctuations. Here, the effect is signalled by 
the appearance of two poles in the fermion propagator: an attempt to inject a 
free fermion into the second
unoccupied energy level above the Fermi surface causes a hole excitation. 
The resulting  hole-electron pair (in the first unoccupied level)
decomposes into spin and charge 
fluctuations which propagate through the medium with different velocities.

This, too reminds of the problem encountered in a Schwinger-Dyson Equation analysis of 
quenched hypercharge \cite{me} where multiple unphysical ``poles'' related 
to several kinds of fermion-antifermion pairing modes were found.
We now see a different possible interpretation of this feature as some kind
of recombination of fermionic degrees of freedom.

\begin{figure}[thb]
\centerline{\rotatebox{270}{\includegraphics{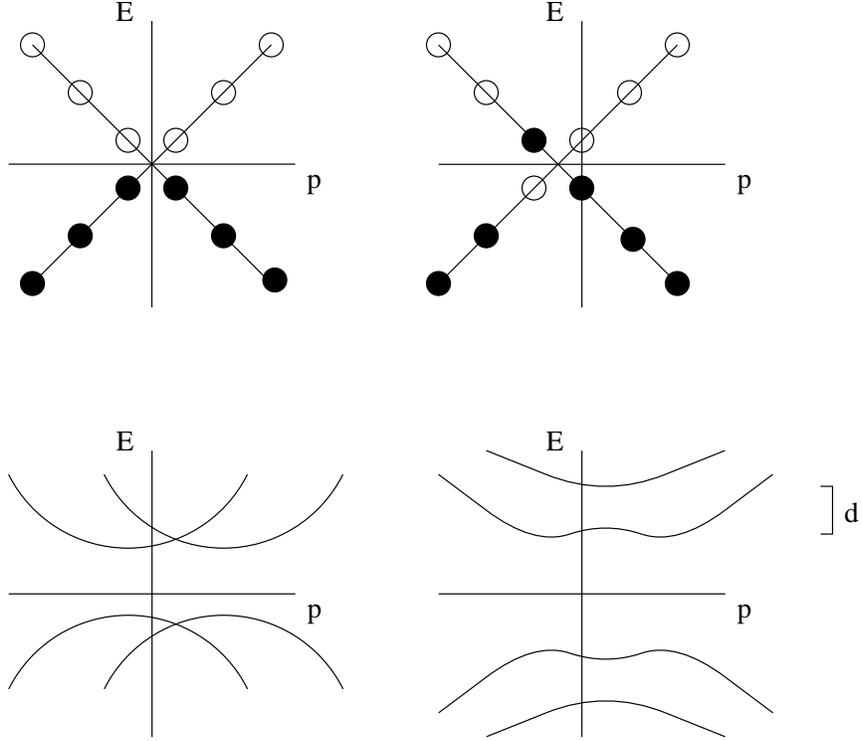}}}
\caption{Dispersion relations for eigenmodes of Eq. (\ref{2d}). 
Black and white circles represent filled and empty states respectively.}
\protect \label{disp}
\end{figure}

A qualitative argument, based on the Dirac sea picture of the anomaly
 \cite{Jackiw} is paraphrased as follows.  
The two-dimensional (for simplicity) hypercharge theory has Lagrangian:
\begin{equation}
{\cal L}_{2D}=\bar{\psi}\gamma^{\mu}(i \partial_{\mu}-(c_{L}\chi_{L}+c_{R}\chi_{R})Z_{\mu}\psi \label{2d}
\end{equation}
In the Dirac sea picture, second quantisation corresponds to filling all
negative energy eigenmodes while leaving positive ones empty.
Setting $Z_{0}=0$ and the potential $Z_{1}=Z$ a space-time constant, the
eigenmodes satisfy the 2-D Dirac equation
\begin{eqnarray*}
E=-\gamma^{\mu}(p_{\mu}-(c_{L}\chi_{L}-c_{R}\chi_{R}))Z_{\mu}
\end{eqnarray*}
For $Z=0$ the energy-momentum dispersion relation 
\begin{eqnarray*}
E=\pm p
\end{eqnarray*}
is shown in the
upper left of Figure \ref{disp}. The left- and right-hand branches correspond
to the separate fermion chiralities. If $Z$ is adiabatically changed to
a small (positive) value the relation is that of the upper right diagram,
\begin{eqnarray*}
E=\left\{ \begin{array}{c} 
p+c_{R}\delta Z,\\
-(p+c_{L} \delta Z), \end{array} \right.
\end{eqnarray*}
where we have assumed both $c_{L}$ and $c_{R}$ are positive.
Gauge transformations in this case cause a nett production of right 
antiparticles and left particles. While the total number of states
is conserved, the separate left and right numbers are not.

Let us now consider an extra ingredient not contained in \cite{Jackiw}:
introducing a mass term $m\bar{\psi}\psi$ to the Lagrangian Eq.(\ref{2d})
would lead to the ``gapped'' lower left diagram in
Fig \ref{disp}. 
\begin{eqnarray*}
E=\pm \left\{\begin{array}{c} 
\sqrt{(p+c_{R}\delta Z)^{2}+m^{2}}\\
\sqrt{(p+c_{L} \delta Z)^{2}+m^{2}} \end{array} \right.
\end{eqnarray*}
Increasing the value $\delta Z$ has the effect of shifting the parabolas
upwards and to the left for positive couplings $c_{R}$, $c_{L}$,
(down and right for negative values)
resulting in a nett production of right-handed particles at the expense of left
ones, as required by a hypothesised dynamical origin for fermion generations \cite{AWT}.
Using this analogy the anomaly-induced collapse of the left 
vacuum required in \cite{AWT} would appear to break down if $c_{L}=-c_{R}$ whereby 
the left and right parabolas shift in opposite directions, such that there
is no net change in chirality. 
This happens for the hypercharge couplings
for the electron, up- and down-type quarks at the values
$\sin^{2} \theta_{W}=1/4$, $3/8$ and $3/4$ respectively.
In this case an alternative scenario, such as a splitting of ``charged'' 
fermion operators into neutral fermions plus
``charged'', `` spinless'' slave bosons, where here the terms in inverted
commas make the analogy with the spin-charge separation in low-dimensional 
systems, could produce fermion generations in a manner outlined below.

Of course these dispersion relations do not correspond to the 
eigenmodes of a full Dirac fermion; each branch contains only half
the required number of degrees of freedom.
Up until now we have considered a single fermion, however the
slave-boson {\it ansatz} is for a many-body phenomenon. If we consider a 
superposition of a fermion and antifermion 
to make up the requisite number of degrees of freedom it is apparent that 
this diagram could also correspond to dispersion relations for two 
distinct bosonic objects, as shown in the lower right picture, in the 
limit of a vanishing gap $d$. These relations, of the form
\begin{eqnarray*}
E^{2}_{1}& = & p^{4} + m^{2}_{1} \\
E^{2}_{2}& = & (p^{2}+ g S)^{2}+(m^{2}_{1}-d)
\end{eqnarray*}
have a natural interpretation as those of composite bosons. $E_{1}$ and
$E_{2}$ would then represent fluctuations in ``charge'' and ``spin''-type 
degrees of freedom respectively.

\section{Standard Model Extensions}

Consider now a free fermion with global $U(2f)_{L}\otimes U(2f)_{R}$
``isoflavour'' symmetry. 
Addition of charge density interactions Eq.(\ref{lut1}) breaks the 
symmetry 
\begin{equation}
U(2f)_{L}\otimes U(2f)_{R}\supset (SU(2f)_{L}\otimes U(1))_{L}\otimes (SU(2f)_{R}\otimes U(1).
\end{equation}
For a single isoflavour, $f=1$, the $U(2)$ fermions decouple into commuting 
$SU(2)\otimes U(1)$ sectors reminiscent of a left-right-symmetric 
electroweak theory.
Upon breaking this to the QED scale, the degeneracy in the (iso)``spin''
condensates is lifted, leading to fermion mass terms of the form 
\begin{eqnarray*}
m=m_{\uparrow}(1+\tau_{3})/2 + m_{\downarrow}(1-\tau_{3})/2.
\end{eqnarray*}
Note that a mass term of this form is used in Gribov's \cite{Gribovgol} calculation of the $W$ 
and $Z$ masses.
The dominant contribution (from the heaviest fermion generation) to vacuum 
polarisation reproduces good approximations to both boson masses and the 
expected Higgs VEV.

Alternatively if instead of Abelian densities of left- and 
right-movers, as in the Luttinger lagrangian, interactions between ``isospin'' 
densities 
\begin{eqnarray*}
\sigma^{a}=\bar{\psi}\tau^{a}\psi, \hspace{0.1cm} a=1,\ldots 3
\end{eqnarray*}
where $\tau^{a}$ are the $SU(2)$ isospin matrices,
the relevant decomposition into decoupled theories is \cite{Andrei}
\begin{equation}
U(2f) \otimes U(2f)\supset [SU(f)_{2} \otimes SU(2)_{f} \otimes U(1)]_{L}\otimes SU(f)_{2}
\otimes SU(2)_{f} \otimes U(1)]_{R} \label{fscs}
\end{equation}
Here the integer subscript denotes the fact that the interaction is in fact 
described by a chiral Wess-Zumino-Witten \cite{WZ,Witten} model, the value 
referring to the central charges.

If we identify $f$ with the number of known fermion generations $f=3$, then 
the model contains not only (iso)spin and (hyper)charge interactions but an 
$SU(3)$ ``flavour'' sector also. 
In two dimensions \cite{Andrei} the analogous model contains a non-trivial
fixed point which generates a mass gap for the fermion propagators.
The bosonic spin fluctuations also acquire mass while the charge and flavour 
excitations remain gapless, strongly reminiscent of photons and gluons
in the SM.

In this context we note the ``dualised''
 standard model 
(DSM) \cite{DSM} also associates the number of fermion generations with that of the fermionic colour
degrees of freedom. This colour ``dual'', analogous to the duality of 
electrodynamics under exchange of charge and magnetism, represents the 
{\it same} gauge symmetry as $SU(3)$, differing only by parity.
The question of whether the decomposition (\ref{fscs}) is equivalent to 
the DSM written in ``left-right'' rather than ``vector-axial'' notation certainly warrants further investigation.

The fact that the centre of the group $SU(3)$ is $Z_{3}$, the permutation group
 of three elements, also serves as
a motivation for the self-consistent introduction of a $Z_{3}$-symmetric 
"fundamental" fermion in the generational model of Kiselev \cite{Kiselev}.
The  $Z_{3}$ would then be interpreted as a relic of the broken dual $SU(3)$.
Moreover a dual relic of the right-handed $SU(2)$ sector of Eq.\ref{fscs}
would be expected. In this context we note the chiral $Z_{3}\otimes Z_{2}$
vacuum symmetry required \cite{me} in a recently proposed modification
to the Kiselev \cite{Kiselev} mechanism.

The model has a natural three-step decomposition, from the ``Luttinger'' phase,
through the left-right symmetric SM, step (\ref{st1}), down to the Standard Model in stage (\ref{st2}) before, finally,
the conventional chiral symmetry breaking reproduces the familiar low-energy physics of stage (\ref{st3}):
\begin{eqnarray}
U(6)_{L}\otimes U(6)_{R} & \supset & [SU(3)_{2}\otimes SU(2) \otimes U(1)]_{L}\otimes
[SU(3)_{2}\otimes SU(2) \otimes U(1)]_{R} \nonumber \\
 & &  \label{st1} \\
& \supset &  SU(3)_{2} \otimes SU(2)_{L} \otimes U(1)_{H}\label{st2} \\
 & \supset & SU(3)_{c} \otimes U(1)_{QED}. \label{st3}
\end{eqnarray}
This symmetry breaking pattern is consistent with the hierarchy of the
three critical chiral scales required for the proposal \cite{AWT}
for a dynamical origin of fermion family structure and masses.
 
In conclusion we observe that certain recent developments in the field of
condensed matter, in particular the behaviour of chiral fluids, are 
potentially fertile ground for understanding how the behaviour of the scalar 
sector of the SM gives rise to fermion mass, generation number and flavour
 mixing.
%
In our opinion there is compelling evidence to investigate whether a 
``technicolour'' version of the left-right symmetric SM might be analogous to
 certain types of slave-boson {\it ans\"{a}tze}. \\

\noindent
\large{{\bf Acknowledgments}}

\noindent
The author wishes to thank A.W. Thomas and G.E. Volovik for helpful comments.

\end{document}